\documentclass[doublecol]{epl2}

\newcommand{\bs}{\boldsymbol}
\newcommand{\beq}{\begin{equation}}
\newcommand{\eeq}{\end{equation}}
\newcommand{\bpm}{\begin{pmatrix}}
\newcommand{\epm}{\end{pmatrix}}
\newcommand{\bea}{\begin{eqnarray}}
\newcommand{\eea}{\end{eqnarray}}
\newcommand{\nnb}{\nonumber}
\newcommand{\la}{\langle}
\newcommand{\ra}{\rangle}

\title{Geometric potentials in quantum optics: \\A semi-classical interpretation}
\shorttitle{Geometric potentials in quantum optics}

\author{M.~Cheneau\inst{1}, S.~P.~Rath\inst{1}, T.~Yefsah\inst{1},
K.~J.~G\"unter\inst{1}, G.~Juzeli\=unas\inst{2}, and J.~Dalibard\inst{1}}
 \shortauthor{M.~Cheneau \etal}

\institute{
  \inst{1} Laboratoire Kastler Brossel and CNRS, Ecole Normale Sup\'erieure,
  24 rue Lhomond, 75005 Paris, France\\
  \inst{2} Institute of Theoretical
Physics and Astronomy of Vilnius University, A. Gostauto 12, Vilnius 01108,
Lithuania}

\abstract{ We propose a semi-classical interpretation of the geometric scalar and
vector potentials that arise due to Berry's phase when an atom moves slowly in a
light field. Starting from the full quantum Hamiltonian, we turn to a classical
description of the atomic centre-of-mass motion while still treating the internal
degrees of freedom as quantum variables. We show that the scalar potential can be
identified as the kinetic energy of an atomic micro-motion caused by quantum
fluctuations of the radiative force, and that the Lorentz-type force appears as a
result of the motion-induced perturbation of the internal atomic state. For a
specific configuration involving two counter-propagating Gaussian laser beams, we
relate the geometric forces to the radiation pressure and dipole forces known from
quantum optics. The simple physical pictures provided by the present analysis may
help for the design and the implementation of novel geometric forces. }

\begin{document}

\maketitle

Cold atomic gases are considered as efficient simulators of quantum condensed matter
systems (for a review, see e.g. \cite{Bloch:2008}). The confinement potential and
the atomic interactions can be tailored almost at will, allowing for example to
mimic with atomic vapours situations encountered for electrons in solid state
materials. A major step in the implementation of these simulators is the possibility
to apply a gauge field to the cold atomic gas in order to model the vector potential
appearing when charged particles are placed in a magnetic field. Up to now such
gauge fields have been mostly obtained by rotating the gas \cite{Fetter:2008}. In
this case the transformation to the rotating frame indeed corresponds to giving the
particles a fictitious charge, and applying an effective uniform magnetic field.
Another method consists in using so-called geometric potentials which can
considerably extend the range of gauge fields realisable in neutral gases. In
particular, they offer the possibility to produce non homogeneous or time dependent
effective orbital magnetism.

Geometric potentials \cite{Mead:1979,Berry:1989,Moody:1989,Bohm:2003} generally
result from Berry's phase \cite{Berry:1984} that appears when particles with an
internal structure move slowly enough that their fast internal dynamics
adiabatically adjusts to the centre-of-mass motion. In quantum optics such geometric
potentials can be generated using laser beams to split the atomic internal energy
levels. This was suggested in \cite{Dum:1996,Visser:1998}, and a first experimental
investigation was presented in \cite{Dutta:1999}. The concept of geometric
potentials can be extended to simulate gauge fields that are more elaborate than the
$U(1)$ symmetry of electromagnetism. Using appropriate laser configurations, one can
in principle implement the general ideas outlined in \cite{Wilczek:1984} to generate
non-Abelian gauge fields \cite{Osterloh:2005,Ruseckas:2005}.

In spite of numerous investigations of possible geometric potentials in quantum
optics, a simple physical interpretation of the forces appearing due to the gauge
fields seems still to be lacking. Of course these forces ultimately arise from the
exchange of momentum between light and atoms, but this generic process can lead to
different physical mechanisms. Since identifying these mechanisms may help to design
future configurations, we propose in this letter a semi-classical analysis of
geometric potentials. Our approach is directly inspired by the formalism used to
calculate the standard radiative forces acting on an atom placed in a laser beam. We
first present the general semi-classical derivation of the two (scalar and vector)
geometric potentials acting on an atom. We then discuss the various physical
pictures that emerge for the concrete implementation that has been proposed in
\cite{Juzeliunas:2006}.

We start with a brief reminder of the standard formalism of geometric potentials, in
which both internal and external (centre-of-mass) atomic degrees of freedom are
treated using quantum mechanics. The relevant internal dynamics is described in an
$N$-dimensional Hilbert space. The atom can be submitted to static electric or
magnetic fields as well as to time-dependent electromagnetic fields. The fields are
supposed to be in a coherent state so that they can be described by classical
functions. Assuming that the time-dependence of the atom-field interaction can be
eliminated using the rotating-wave approximation, the eigenstates of the atom-field
coupling form an orthogonal basis $\{|\psi_j(\bs r)\ra, \ j=1,\ldots, N\}$ of the
internal Hilbert space of the atom at any point $\bs r$. We denote $E_j(\bs r)$ the
corresponding energies. The Hamiltonian of the problem is thus
 \bea
\hat H &=& \frac{\hat {\bs p}^2}{2M}+\hat V(\bs r)\ ,\\
\hat V(\bs r) &=& \sum_{j=1}^N E_j(\bs r)\;\hat Q_j(\bs r)\ . \label{eq:coupling}
 \eea
Here, $M$ is the atomic mass, $\hat {\bs p}=-i\hbar \bs \nabla$ the centre-of-mass
momentum operator and $\hat Q_j(\bs r)=|\psi_j(\bs r)\rangle \langle \psi_j(\bs r)|$
the projector onto the $j$-th internal eigenstate. Suppose now that the energy of
one of the internal eigenstates, say $|\psi_1\ra$, is well separated from the other
ones. We choose the initial internal atomic state equal to $|\psi_1(\bs r)\ra$ at
any point $\bs r$ and suppose that the atom moves slowly enough for the adiabatic
theorem to hold. The internal state then remains equal to $|\psi_1\ra$ at any time,
and the energy $E_1(\bs r)$ plays the role of a potential energy for the
centre-of-mass motion. In addition, the geometric phase accumulated by the atom
gives rise to additional vector $\bs A(\bs r)$ and scalar $U(\bs r)$ potentials so
that the atom Hamiltonian in the adiabatic approximation
\cite{Mead:1979,Berry:1989,Moody:1989,Bohm:2003} reads
 \beq
\hat H_{\rm adiab.}= \frac{\left(\hat {\bs p}- {\bs A}(\bs r)\right)^2}{2M}+E_1(\bs
r)+ U(\bs r) \ ,
 \label{eq:Hadiabatic}
 \eeq
with
 \bea
{\bs A}({\bs r})&=& i\hbar\; \la \psi_1(\bs r) |\bs \nabla\psi_1(\bs r)\ra
\ , \label{eq:vector}\\
U(\bs r)&=&\frac{\hbar^2}{2M}\sum_{j\neq 1} \left|
 \la \psi_1(\bs r) | \bs \nabla\psi_j(\bs r)\ra
\right|^2\ , \label{eq:scalar}
 \eea
where we note by convention $|\bs \nabla \psi(\bs r)\ra=\bs \nabla \left(|\psi(\bs
r)\ra\right)$.

The goal of this letter is to provide a simple physical interpretation of these
potentials within the framework of a \emph{semi-classical} analysis. Here, the term
\emph{semi-classical} refers to the fact that we describe the atomic internal
degrees of freedom using quantum mechanics, but we treat classically the
centre-of-mass motion. Within this approximation we want to recover the equation of
motion
 \beq
M \frac{d\bs v}{dt}=  -\bs \nabla E_1 (\bs r)
 -\bs \nabla  U(\bs r)+\bs v \times \bs B(\bs r)
 \label{eq:motion}
 \eeq
corresponding to the Hamiltonian (\ref{eq:Hadiabatic}). This equation of motion
involves three forces. The first one is simply the gradient of the energy $E_1$ of
the occupied level. The second one originates from the scalar potential $U(\bs r)$,
and the third one has the structure of a Lorentz force for a charge $q=1$ in an
effective magnetic field $\bs B(\bs r)= \bs \nabla \times \bs A(\bs r)$.

The main tool for the semi-classical analysis is the force operator acting on the
$N$-dimensional Hilbert space:
 \beq
 \hat {\bs F}(\bs r)=-\bs \nabla {\hat V}=-\sum_j \left( \bs \nabla(E_j)\; \hat Q_j + E_j \; \bs
 \nabla (\hat Q_j) \right).
 \label{eq:forceop}
 \eeq
Knowing the internal state $|\phi\ra$  for an atom at point $\bs r$ (a concept which
is valid within the semi-classical approach), we will be able to calculate the
average force $\la \phi |\hat {\bs F}(\bs r)|\phi\ra$ and the correlation functions
of the force operator. This semi-classical approach has been very fruitful for the
study of the radiative forces acting on an atom irradiated by laser beams
\cite{Gordon:1980}, and its connection with a full quantum description of the atomic
motion in laser light is well established \cite{Dalibard:1985}.

\section{Origin of the scalar potential}

In this section we consider an atom with zero centre-of-mass velocity, so that the
Lorentz force in (\ref{eq:motion}) is also zero. The atom internal state is supposed
to be $|\phi\ra=|\psi_1\ra$ and we immediately get from (\ref{eq:forceop})
 \beq
\la \hat{\bs F}(\bs r)\ra=-\bs \nabla E_1\ ,
 \eeq
where we used that the states $\{|\psi_j(\bs r)\ra\}$ form a normalised orthogonal
basis at any point $\bs r$. We thus recover the first term in eq.~(\ref{eq:motion})
but not the force originating from the scalar potential $U(\bs r)$. This could be
expected since the expression (\ref{eq:scalar}) of the scalar potential involves the
atomic mass $M$, which does not enter in the expression (\ref{eq:forceop}) of the
force operator $\hat {\bs F}$. The scalar potential can still be recovered within
the semi-classical approach, as we show now, provided one goes one step beyond the
mere calculation of the average force.

The starting point of our reasoning consists in noting that the state $|\psi_1\ra$
occupied by the atom is an eigenstate of the coupling Hamiltonian $\hat V$, but
\emph{not} an eigenstate of the force operator $\hat{\bs F}$. Hence, $\langle \hat
{\bs F}^2 \rangle \neq \la \hat {\bs F}\ra ^2$ or, in physical terms, the force
acting on the atom undergoes quantum fluctuations around its average value. As we
will see, these fluctuations occur at the typical Bohr frequencies
$\omega_{j1}=(E_j-E_1)/\hbar$ of the internal atomic motion. For the adiabatic
approximation to hold, these frequencies have to be much larger than the
characteristic frequencies of the external atomic motion. Consequently, the quantum
fluctuations of the force cause a fast micro-motion of the atom, similar to the one
arising for charged particles in a Paul trap \cite{Paul:1990}. The kinetic energy of
the micro-motion then plays the role of a potential for the slow motion of the
atomic centre-of-mass \cite{Landau:Mechanics}. We demonstrate below that this
kinetic energy coincides with the scalar potential $U(\bs r)$.

The fluctuations of the force operator $\hat{\bs F}$ are characterised by the
symmetrised correlation function of the operator $\delta \hat{\bs F }=\hat{\bs
F}-\la \hat{\bs F}\ra$, calculated in the Heisenberg picture:
 \beq
C(t,t')=\frac{1}{2}\la \delta \hat{\bs F }(t)\cdot \delta \hat{\bs F }(t')
 \;+\; \delta \hat{\bs F }(t')\cdot \delta \hat{\bs F }(t)
\ra \ .
 \eeq
Since the average is taken in an eigenstate of the Hamiltonian, $C(t,t')$ depends
only on the time difference $\tau=t-t'$ and we obtain after some algebra
 \beq
C(\tau)=\sum_{j\neq 1} C_j \cos\left( \omega_{j1}\tau \right)\ ,
 \eeq
with
 \beq
 C_j=\hbar^2\omega_{j1}^2 \;|\la \psi_1 |\bs \nabla \psi_j\ra|^2\ .
 \label{eq:Aj}
 \eeq
In order to understand the consequences of these fluctuations, consider a classical
particle submitted to a stochastic force $\bs F(t)$ such that $ \overline {\bs
F(t)}=0$ and $\overline {\bs F(t)\cdot \bs F(t')}=C(t-t')$. The Fourier transform $
\bs f(\omega)$ of $\bs F(t)$ is such that $\overline{ \bs f(\omega) }=0$ and
$\overline{ \bs f^*(\omega)\cdot \bs f(\omega')}=\delta(\omega-\omega')\;B(\omega)$,
where
 \beq
B(\omega)=\frac{1}{2}\sum_{j\neq 1} C_j \left( \delta(\omega+\omega_{j1})+
\delta(\omega-\omega_{j1})\right)
 \eeq
is the Fourier transform of $C(\tau)$. The solution of the equation of motion $\dot
{\bs p}=\bs F$ is
 \beq
\bs p(t)=\int \frac{\bs f(\omega)}{i\omega}\;e^{i\omega t}\;d\omega
 \eeq
and has a zero average. However, the average kinetic energy is strictly positive and
equal to
 \beq
\frac{\overline{\bs p^2}}{2M}=\int \frac{B(\omega)}{2M\omega^2}\;d\omega\ .
 \label{eq:kinetic}
 \eeq
In the explicit calculation of (\ref{eq:kinetic}), the contribution of the
$\omega^{-2}$ denominator cancels out the transition frequencies $\omega_{j1}$ that
appear in the expression (\ref{eq:Aj}) of $C_j$. Finally one exactly recovers the
result (\ref{eq:scalar}) for the scalar potential. This validates the interpretation
of this potential in terms of the kinetic energy of the atomic micro-motion.

The above result sheds new light on the Hamiltonian (\ref{eq:Hadiabatic}) of the
full quantum description. We can now interpret the term $M\hat{\bs v}^2/2$, with
$M\hat{\bs v}=\hat{\bs p}-\bs A$, as the kinetic energy of the slow centre-of-mass
motion, whereas the kinetic energy of the fast micro-motion builds the scalar
potential $U(\bs r)$. It is also interesting to connect the present analysis of the
scalar geometric potential with the intriguing problem of a two-level atom moving
around the node of a standing wave. In the latter case, it is found that the average
force acting on the atom is zero, as expected since the light intensity vanishes at
the nodes. However, the atomic momentum diffusion coefficient, which is also related
to the correlation function of the force operator, is non zero
\cite{Gordon:1980,Cohen-Tannoudji:1992}.

\section{Origin of the Lorentz force}

We now consider the case of a slowly moving atom and calculate the average of the
force operator (\ref{eq:forceop}) at first order in velocity. More precisely, we
assume an atom initially at rest in the internal state $|\phi\ra=|\psi_1\ra$, that
is adiabatically set in motion to reach a velocity $\bs v$. Because of this motion
the internal atomic state contains some admixture of the other eigenstates
$|\psi_j\ra$ and the average force is different from the zero-velocity result.

We write the internal state as $|\phi\ra=\sum_j \alpha_j |\psi_j\ra$ and solve the
Schr\"odinger equation as a power series in velocity. The procedure detailed in the
appendix gives the coefficients $\alpha_j$ at first order in $v$:
 \beq
\alpha_j(t)\simeq i\hbar \frac{\bs v \cdot \la \psi_j |\bs \nabla \psi_1\ra
}{E_j-E_1}\,e^{-iE_1t/\hbar}\qquad (j\neq 1)\ .
 \label{eq:perturb}
 \eeq
The calculation of the average force at first order in $v$ is also outlined in the
appendix and leads to
 \beq
\la \hat {\bs F}\ra=i\hbar
   \la \bs \nabla \psi_1|\, \left( \bs v \cdot |\bs \nabla  \psi_1\ra
  \right)
\ +\ \mbox{c.c.}
 \label{eq:Lorentz}
 \eeq
One can readily check that this expression coincides with the Lorentz force $\bs
v\times \bs B$ appearing in eq.~(\ref{eq:motion}).

This way of recovering the Lorentz force is very reminiscent of the general
semi-classical calculation of the velocity-dependent radiative forces
\cite{Gordon:1980,Dalibard:1985}. There is one important difference, however. In the
latter case one generally finds $\bs F \cdot \bs v\neq 0$, which is at the origin of
laser cooling (for example via the Doppler or Sisyphus mechanisms). In the
particular case considered here, no photon spontaneous emission process occurs,
dissipation is absent, and we are left with a Lorentz force of geometric origin.

\section{Illustration for a particular atom-laser configuration}

\begin{figure}
\centerline{\includegraphics{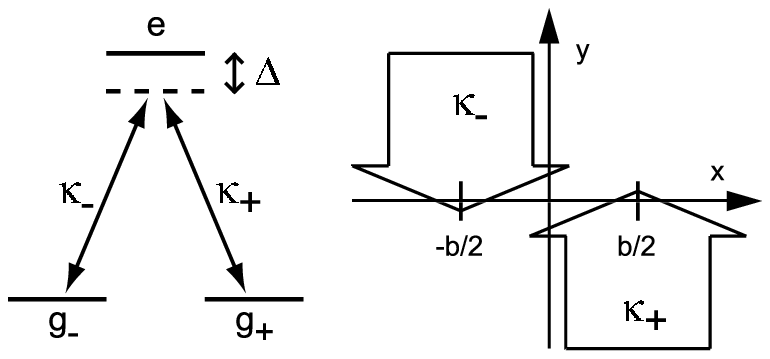}} \caption{Atomic level scheme and laser
configuration proposed in \cite{Juzeliunas:2006} to generate a vector and a scalar
potential.}
 \label{fig.1}
\end{figure}

We now relate the geometric forces in quantum optics to the known radiative forces
that generally act on an atom irradiated by one or several laser beams. We thus
explain the mechanisms at the origin of the geometric potentials in terms of
exchange of momentum between atoms and light. For this purpose, we turn to the
specific configuration sketched in fig. \ref{fig.1}, which has been proposed in
\cite{Juzeliunas:2006}. A three-level atom with two degenerate ground states
$|g_{\pm}\ra$ and an excited state $|e\ra$ is irradiated by two counter-propagating
laser beams. The beam propagating in the $+y$ (resp.~$-y$) direction drives the
transition $g_+ \leftrightarrow e$ (resp.~$g_- \leftrightarrow e$). The two beams
have the same waist $w$ and the same intensity. They are offset with respect to the
$y$ axis by a distance $\pm b/2$, where $b$ is typically of the order of $w$.

Using the rotating-wave approximation the atom-light coupling can be written as
 \beq
\hat V(\bs r)=-\hbar \Delta |e\ra \la e| + \sum_{j=\pm} \left(\hbar \kappa_j(\bs r)
|e\ra \la g_j| +\mbox{h.c.}\right)\ .
 \eeq
The Rabi frequencies $\kappa_\pm$ are given by
 \beq
\kappa_\pm (\bs r)= \kappa\; e^{\pm iky}\;e^{-(x\mp b/2)^2/w^2}\ ,
 \eeq
where $k$ is the modulus of the wave vector of the laser beams. We assume that the
Raman resonance condition is satisfied between the two ground levels $g_\pm$, and we
denote $\Delta$ the detuning of the laser frequency with respect to the $g_\pm
\leftrightarrow e$ transition.

A well known characteristics of this configuration is that one of the three
eigenstates of $\hat V(\bs r)$ involves only the two ground states $|g_\pm\ra$ and
has zero overlap with the excited state $|e\ra$
\cite{Arimondo:1996,Lukin:2003,Fleischhauer:2005}. An atom prepared in this
so-called \emph{non-coupled} or \emph{dark} state will not undergo any spontaneous
emission process, which is a key feature for the practical use of geometric
potentials. The non-coupled state is
 \beq
| \psi_1\ra = \frac{\kappa_-}{\Omega}|g_+\ra\;-\; \frac{\kappa_+}{\Omega}|g_-\ra\ ,
 \label{eq:noncoup}
 \eeq
where we set $\Omega=(|\kappa_-|^2+|\kappa_+|^2)^{1/2}$. The corresponding energy at
any point is $E_1(\bs r)=0$ so that the first term on the right hand side of
(\ref{eq:motion}) vanishes. The scalar potential and the effective magnetic field
are \cite{Juzeliunas:2006}:
 \bea
 U(x)&=&\frac{\hbar^2 (k^2+b^2/w^4)}{2M}\; G(x) \label{eq:Ulambda}\ ,\\
\bs B(x)&=& -\frac{2 \hbar k b}{w^2}\;G(x)\;\bs u_z\ ,
 \label{eq:Blambda}
 \eea
where $G(x)=\cosh^{-2}(2xb/w^2)$.

We start our analysis with the scalar potential $U(\bs r)$. When $x\to \pm \infty$,
it tends to zero as expected from its interpretation as the energy of a
micro-motion, which vanishes when the atom sits outside the laser beams. The value
of $U$ at $\bs r=0$ can also be recovered easily. At this point the two components
of the force operator are
 \bea
\hat F_x &=& -F_0\;\left( |e\ra \la g_a| + |g_a\ra \la e| \right)\ ,
 \label{eq:Fx}\\
\hat F_y&=& -iF_1 \left( |e\ra \la g_a|-|g_a\ra \la e|\right)\ ,
 \label{eq:Fy}
 \eea
where $F_0=\sqrt 2\,\hbar \kappa b/w^2$, $F_1=\sqrt 2\, \hbar k \kappa$ and where we
have introduced the antisymmetric combination $|g_{a}\ra= (|g_+\ra-|g_-\ra)/\sqrt
2$. At $\bs r=0$, $|\psi_1\ra=|g_a\ra$ so that $\la \hat F_{x,y}\ra=0$, which
directly follows from $E_1(\bs r)=0$. The correlation function of the force is
readily calculated; choosing $\Delta=0$ for simplicity, we find
$C(\tau)=(F_0^2+F_1^2)\cos(\sqrt2\, \kappa \tau)$. The kinetic energy of the
micro-motion induced by this oscillating force is $(F_0^2+F_1^2)/(4M\kappa^2)$ and
indeed coincides with the general result (\ref{eq:Ulambda}) at $x=0$. For realistic
laser configurations the waist $w$ and the offset $b$ are large compared to $k^{-1}$
so that the dominating contribution is due to $F_1$, corresponding to a micro-motion
directed along the propagation axis $y$ of the beams.

The two operators $\hat F_x$ and $\hat F_y$ given in (\ref{eq:Fx}-\ref{eq:Fy}) are
the so-called \emph{dipole force} operator and \emph{radiation pressure force}
operator, respectively. The dipole force $\hat F_x$ originates from the intensity
gradient of the laser beams along the $x$ axis. In terms of photon momentum
exchange, it can be understood as a redistribution of photons between the various
plane waves that contribute to the formation of the intensity gradient (see e.g.
\cite{Cohen-Tannoudji:1992}). The radiation pressure force $\hat F_y$ originates
from the phase variation $e^{\pm iky}$ of the laser beams along the $y$-axis. It
leads to changes of atomic momentum by $\pm \hbar k$ when the atom absorbs a photon
from one of the laser beams. Interestingly, the cancellation of the average force
acting on an atom at rest at $\bs r=0$ in the internal state $|g_a\ra$ results from
a destructive interference. For example, $\la \hat F_x\ra=0$ because the state
$|g_a\ra$  is a combination with equal weights of the two eigenvectors
$|\chi_\pm\ra=(|g_a\ra \pm |e\ra)/\sqrt 2$ of the force operator $\hat F_x$. The
dipole force felt by an atom in state $|\chi_\pm\ra$ is $\mp F_0$, so that the
average force vanishes for an atom prepared in $|g_a\rangle$.

We now turn to the case of a moving atom and to the interpretation of the Lorentz
force. As the force is linear in velocity, we can study separately the cases of a
motion parallel to the $x$ and to the $y$ axis. We first consider an atom moving
along the $y$ axis with velocity $v_y$. Its internal state differs from the value
$|g_a\ra$ of an atom at rest, and its expression at first order in $v_y$, as deduced
from (\ref{eq:perturb}), is
 \beq
|\phi\ra= |g_a\ra + \frac{kv_y}{\sqrt 2\, \kappa}|e\ra\ ,
 \label{eq:rotated}
 \eeq
where we have again chosen $\Delta=0$ for simplicity. This state can also be
calculated directly by switching to the atom rest frame, which amounts to adding the
small perturbation $ \delta\hat V= kv_y(\hat Q_--\hat Q_+)$ to the atom-light
coupling, where $\hat Q_\pm$ is the projector on the state $|g_\pm\ra$. One can
check that the state $|\phi\ra$ is an eigenstate of the perturbed coupling $\hat
V+\delta \hat V$ with the same eigenvalue 0. The average force acting on an atom in
the state (\ref{eq:rotated}) is not zero anymore. The perfect balance between the
two dipole forces $\pm F_0$ is now broken to the benefit of $-F_0$ ($+F_0$) if
$v_y>0$ ($v_y<0$). The explicit calculation of $\la \phi|\hat F_x|\phi\ra$ is
immediate and yields exactly the Lorentz force value $-2\hbar kb v_y/w^2$ obtained
from eq.~(\ref{eq:Blambda}). In the case of a motion along the $y$ axis, the Lorentz
force is therefore a direct consequence of the dipole potential. Remarkably, the
Lorentz force is independent of the Rabi frequency $\kappa$ although the dipole
force amplitude $F_0$ is proportional to $\kappa$. The reason is that the
``rotation" angle of the state $|\phi\ra$ with respect to the non-coupled state
scales as $\kappa^{-1}$ [see eq.~(\ref{eq:rotated})] and this scaling exactly
compensates for the $\kappa$ dependence of $F_0$.

We finally consider an atomic motion along the $x$ axis, with a Lorentz force along
the $y$-axis, originating thus from the radiation pressure force operator $\hat
F_y$. This case could be analysed along the lines of eqs.
(\ref{eq:perturb}-\ref{eq:Lorentz}), but it is actually more instructive to look at
the cumulative effect of the Lorentz force, i.e. the change of momentum along $y$,
when the atom crosses the two laser beams. Suppose that the atom is initially
located at time $t_1$ at a point $x_1<0$ with $|x_1|\gg w$. At this location
$\kappa_+/\kappa_-=\exp(2x_1b/w^2)\ll 1$, so that eq.~(\ref{eq:noncoup}) leads to
$|\psi_1\ra \simeq |g_+\ra$. At a later time $t_2$ the atom has reached the point
$x_2=-x_1$, where $|\psi_1\ra \simeq |g_-\ra$. The momentum change caused by the
Lorentz force is aligned with the $y$ axis and its value is
 \bea
\Delta p_y &=& -\int_{t_1}^{t_2} B_z v_x\, dt=-\int_{x_1}^{x_2}B_z(x)\,dx \nnb \\
&=& \hbar k \left[\tanh(2x_2b/w^2)-\tanh(2x_1b/w^2)\right]\ .
 \eea
With the above assumption that $|x_1|=|x_2|\gg w$ and $b\sim w$, we obtain
 \beq
\Delta p_y \simeq 2 \hbar k\ .
 \eeq
This expression has a clear physical interpretation in terms of photon absorption
and stimulated emission. When the atom moves from $x_1$ to $x_2$, its internal state
rotates adiabatically from $|g_+\ra$ to $|g_-\ra$. This happens by the absorption of
a photon from the wave propagating in the $+y$ direction, driving the $g_+\to e$
transition, and the stimulated emission of a photon in the wave propagating in the
$-y$ direction, driving the $e\to g_-$ transition. The origin of the Lorentz force
in this case is therefore closely related to the so-called STIRAP process
(Stimulated Raman Adiabatic Passage) \cite{Bergmann:1998}.

\section{Discussion and conclusion}

We have given here a semi-classical interpretation of the geometric forces that act
on a slowly moving particle with multiple, spatially varying, internal energy
levels. Such a situation is encountered in quantum optics when an atom moves in the
light field created by several quasi-resonant laser beams. The main assumption is
that the atom follows adiabatically one of its internal energy levels which is
supposed to be well separated from all the other ones. Under these conditions the
atomic centre-of-mass motion involves a scalar and a vector potential. The scalar
potential originates from the kinetic energy of the micro-motion of the atom, under
the action of the quantum fluctuations of the radiative force operator. The
Lorentz-type force associated with the vector potential results from the
perturbation of the atomic internal state due to the slow atomic motion. This
changes the expectation value of the force with respect to an atom at rest, but does
not induce any dissipation. The way we obtain our results is very reminiscent of the
general derivation of the radiative forces created by quasi-resonant laser beams.
Taking as an example the configuration proposed in \cite{Juzeliunas:2006}, we have
related the intervening geometric forces with either the radiation pressure force or
the dipole force.

Our approach can be viewed as a quantum version (for the internal degrees of
freedom) of the fully classical results of \cite{Aharonov:1992}, where the authors
analysed the motion of a particle with a permanent magnetic moment placed in a
strong, non homogeneous magnetic field. There, a rapid oscillation of the magnetic
moment superimposed onto a slow secular motion was found. This rapid oscillation
gives rise to a scalar potential similar to the one of interest here. When the
particle with its magnetic moment initially aligned with the magnetic field was set
in motion, it was found that the magnetic moment acquires a non-zero component in a
direction perpendicular to the magnetic field. The Lorentz-type force that emerged
in this situation is also very similar to the present one.

To conclude, geometric potentials may constitute in the future a key ingredient for
the realisation of a general `quantum gas toolbox' that allows one to address
various open problems of many-body physics. We hope that the simple physical
pictures provided by the present analysis may help for the design and the
implementation of novel geometric forces in this context.

\section{Appendix: Derivation of the Lorentz force}

In this appendix we detail the calculation of the perturbed state $|\phi\ra$ of a
slowly moving atom and the resulting force $\la \hat {\bs F}\ra$. At time $0$ the
atom is at rest in the internal state $|\phi\ra=|\psi_1\ra$. We suppose that it is
uniformly accelerated during a time $T$ to a velocity $\bs v$: $\bs v(t)=\bs v\,
t/T$ for $0\leq t\leq T$. Our aim is to calculate $|\phi\ra$ and $\la \hat {\bs
F}\ra$ at time $T$.

The general expression of the internal atomic state is $|\phi(t)\ra=\sum_j
\alpha_j(t) |\psi_j(\bs r(t))\ra$, leading to the average of the force operator
(\ref{eq:forceop}):
 \bea
\la \hat {\bs F}\ra &=& - \sum_j \bs \nabla E_j \; |\alpha_j|^2 \nnb \\
&& +\sum_{j,k} (E_k-E_j)\;\alpha_k^* \alpha_j\; \la \psi_k |\bs \nabla\psi_j\ra \ .
 \label{eq:forceappen}
 \eea
To calculate the force at first order in $v$, we need all coefficients $\alpha_j$
also at first order. Using the Schr\"odinger equation $i\hbar |\dot \phi\ra=\hat
V(\bs r(t))|\phi\ra$ and $ |\dot \phi\ra=\sum_j \dot \alpha_j |\psi_j\ra + \alpha_j
\,\bs v\cdot |\bs \nabla \psi_j\ra $, we get the corresponding equations of motion:
 \beq
\dot\alpha_j = -iE_j\alpha_j/\hbar +\sum_k \alpha_k \, \bs v \cdot \la \psi_j|\bs
\nabla \psi_k\ra\ .
 \eeq
At order zero all $\alpha_j$'s are zero except $\alpha_1(t)=\exp(-iE_1t/\hbar)$. At
order one we obtain for $j\neq 1$
 \beq
\alpha_j(T)=-\bs v \cdot \la \psi_j|\bs \nabla \psi_1\ra \; e^{-iE_jT/\hbar}\int_0^T
e^{i(E_j-E_1)t/\hbar}\; \frac{t}{T} \; dt \ ,
 \eeq
where $|\psi_j\ra$, $|\psi_1\ra$, $E_j$ and $E_1$ are taken at order zero in $v$,
hence at the location $\bs r$ of the atom at time $T$. Assuming that the atom is
adiabatically set into motion, \emph{i.e.} $T(E_j-E_1)/\hbar \gg 1$, we get
 \beq
\alpha_j(T)\simeq i\hbar \frac{\bs v \cdot \la \psi_j |\bs \nabla \psi_1\ra
}{E_j-E_1}\,e^{-iE_1T/\hbar}\ .
 \label{eq:perturbapp}
 \eeq
At order one in $v$ the equation of motion for $\alpha_1$ is
 \beq
\dot \alpha_1=-i (E_1-\bs v\cdot \bs A)\,\alpha_1/\hbar\ ,
 \label{eq:alpha1app}
 \eeq
whose solution is a number of modulus 1. The two results (\ref{eq:perturbapp}) and
(\ref{eq:alpha1app}) entail that the first part of $\la \hat {\bs F}\ra$ in
(\ref{eq:forceappen}) has no first order component in $v$ since the contributions of
the $\alpha_j$'s for $j\neq 1$ are at least of order 2, and the contribution of
$\alpha_1$ is independent of $v$. In the second part of (\ref{eq:forceappen}), the
only relevant terms are those where one of the two indices $k$ or $j$ equals 1.
Applying the closure relation and keeping the terms linear in velocity, we finally
reach the result (\ref{eq:Lorentz}) for the average force. We note that our
procedure is similar to the original derivation of the geometric phase
\cite{Berry:1984}, which emerges from the term $\bs v\cdot \bs A$ in
(\ref{eq:alpha1app}).

\acknowledgments

We are grateful to Claude Cohen-Tannoudji, Fabrice Gerbier and Ady Stern for useful
discussions. K.~J.~G. and S.~P.~R. acknowledge support from EU (contract IEF/219636)
and from the German Academic Exchange Service (DAAD, grant D/06/41156),
respectively. This work is supported by R\'egion \^Ile de France (IFRAF), CNRS, the
French Ministry of Research, ANR (Project Gascor, NT05-2-42103) and the EU project
SCALA. Laboratoire Kastler Brossel is a \emph{Unit\'e mixte de recherche} of CNRS,
Ecole Normale Sup\'erieure and Universit\'e Pierre et Marie Curie.

\bibliographystyle{eplbib}

\begin{thebibliography}{10}
\expandafter\ifx\csname url\endcsname\relax\def\url#1{\texttt{#1}}\fi

\bibitem{Bloch:2008}
\Name{Bloch I., Dalibard J. \and Zwerger W.} \REVIEW{Rev. Mod.
Phys.}{80}{2008}{885}.

\bibitem{Fetter:2008}
\Name{Fetter A. L.} \REVIEW{arXiv:0801.2952}{}{2008}{}

\bibitem{Mead:1979}
\Name{Mead C.~A. \and Truhlar D.~G.} \REVIEW{J. Chem. Phys.}{70}{1979}{2284}.

\bibitem{Berry:1989}
\Name{Berry M.~V.} \Book{The quantum phase, five years after} in
  \Book{Geometric {P}hases in {P}hysics}, edited by \Name{Shapere A. \and
  Wilczek F.} (World Scientific, Singapore) 1989 pp. 7--28.

\bibitem{Moody:1989}
\Name{Moody J., Shapere A. \and Wilczek F.} \Book{Adiabatic {E}ffective
  {L}agrangians} in \Book{Geometric {P}hases in {P}hysics}, edited by
  \Name{Shapere A. \and Wilczek F.} (World Scientific, Singapore) 1989 pp.
  160--183.

\bibitem{Bohm:2003}
\Name{Bohm A., Mostafazadeh A., Koizumi H., Niu Q. \and Zwanziger J.} \Book{The
{G}eometric {P}hase in {Q}uantum {S}ystems}
  (Springer, Berlin Heidelberg New York) 2003.

\bibitem{Berry:1984}
\Name{Berry M.~V.} \REVIEW{Proc. R. Soc. Lond. A}{392}{1984}{45}.

\bibitem{Dum:1996}
\Name{Dum R. \and Olshanii M.} \REVIEW{Phys. Rev. Lett.}{76}{1996}{1788}.

\bibitem{Visser:1998}
\Name{Visser P.~M. \and Nienhuis G.} \REVIEW{Phys. Rev. A}{57}{1998}{4581}.

\bibitem{Dutta:1999}
\Name{Dutta S.~K., Teo B.~K. \and Raithel G.} \REVIEW{Phys. Rev.
Lett.}{83}{1999}{1934}.

\bibitem{Wilczek:1984}
\Name{Wilczek F. \and Zee A.} \REVIEW{Phys. Rev. Lett.}{52}{1984}{2111}.

\bibitem{Osterloh:2005}
\Name{Osterloh K., Baig M., Santos L., Zoller P. \and Lewenstein M.}
  \REVIEW{Phys. Rev. Lett.}{95}{2005}{010403}.

\bibitem{Ruseckas:2005}
\Name{Ruseckas J., Juzeli\=unas G., \"Ohberg P. \and Fleischhauer M.}
  \REVIEW{Phys. Rev. Lett.}{95}{2005}{010404}.

\bibitem{Juzeliunas:2006}
\Name{Juzeli\=unas G., Ruseckas J., \"Ohberg P. \and Fleischhauer M.}
  \REVIEW{Phys. Rev. A}{73}{2006}{025602}.

\bibitem{Gordon:1980}
\Name{Gordon J.~P. \and Ashkin A.} \REVIEW{Phys. Rev. A}{21}{1980}{1606}.

\bibitem{Dalibard:1985}
\Name{Dalibard J. \and Cohen-Tannoudji C.} \REVIEW{J. Phys. B: At. Mol.
Phys.}{18}{1985}{1661}.

\bibitem{Paul:1990}
\Name{Paul W.} \REVIEW{Rev. Mod. Phys.}{62}{1990}{531}.

\bibitem{Landau:Mechanics}
\Name{Landau L.~D. \and Lifshitz E.~M.} \Book{Mechanics}
  (Butterworth-Heinemann; 3rd edition, Oxford) 1982.

\bibitem{Cohen-Tannoudji:1992}
\Name{Cohen-Tannoudji C.} \Book{Atomic motion in laser light} in
  \Book{Fundamental {S}ystems in {Q}uantum {O}ptics}, edited by \Name{Dalibard
  J., Raimond J.-M. \and Zinn-Justin J.} (North-Holland Amsterdam) 1992 pp.
  1--164.

\bibitem{Arimondo:1996}
\Name{Arimondo E.} \Book{Coherent population trapping in laser spectroscopy} in
  \Book{Progress in Optics, vol. 35}, edited by \Name{Wolf E.} (Elsevier) 1996
  p. 259.

\bibitem{Lukin:2003}
\Name{Lukin M.~D.} \REVIEW{Rev. Mod. Phys.}{75}{2003}{457}.

\bibitem{Fleischhauer:2005}
\Name{Fleischhauer M., Imamoglu A. \and Marangos J.~P.} \REVIEW{Rev. Mod.
Phys.}{77}{2005}{633}.

\bibitem{Bergmann:1998}
\Name{Bergmann K., Theuer H. \and Shore B.~W.} \REVIEW{Rev. Mod.
Phys.}{70}{1998}{1003}.

\bibitem{Aharonov:1992}
\Name{Aharonov Y. \and Stern A.} \REVIEW{Phys. Rev. Lett.}{69}{1992}{3593}.

\end{thebibliography}

\end{document}